\documentclass[11pt]{article}

\usepackage{epsfig}
\usepackage{amsmath}

\providecommand{\rvec}{\mbox{\boldmath$r$}}
\providecommand{\pvec}{\mbox{\boldmath$p$}}
\providecommand{\xvec}{\mbox{\boldmath$x$}}
\providecommand{\rpot}{U_{\textrm{\tiny R}}}
\providecommand{\mass}{\overline{m}^2_0}
\providecommand{\lapl}{a^2\nabla_{\textrm{\tiny L}}^2}
\providecommand{\gfo}{g_{\textrm{\tiny FO}}}

\begin{document}


\renewcommand{\thefootnote}{\fnsymbol{footnote}}

\begin{center}
\textbf{\large Phase structure of self-gravitating systems} \\
\vspace{0.5truecm}
Victor Laliena \\
\vspace{0.25truecm}
{\footnotesize
\textit{Departamento de F\'{\i}sica Te\'orica, Universidad de
Zaragoza, \\
Pedro Cerbuna 12, E-50009 Zaragoza (Spain) 
\\
\vspace{0.15truecm}
laliena@unizar.es \\
}}
\vspace{0.25truecm}
January 16, 2003
\end{center}

\vspace{1truecm}

\begin{center}
\textbf{Abstract}
\end{center}

\noindent
The equilibrium properties of classical self-gravitating systems in
the grand canonical ensemble are studied by
using the correspondence with an euclidean field theory with 
infrared and ultraviolet cutoffs.
It is shown that the system
develops a first order phase transition between a low and a high
density regime. In addition, due to the long range of the 
gravitational potential, the system is close to criticality within
each phase, with the exponents of mean field theory. 
The coexistence of a sharp first order transition and critical
behavior can explain both the presence of voids in large regions of the
universe as well as the self-similar density correlations 
in terms of self-gravity alone.

\vspace{0.5truecm}

{\footnotesize
\noindent
\textit{Keywords}: Field Theory; Critical Phenomena; Long Range Forces \\
\vspace{0.15truecm}
\noindent
PACS: 11.10.-z, 64.60.-i, 05.20.-y
}

\vfill\eject


\renewcommand{\thefootnote}{\arabic{footnote}}
\setcounter{footnote}{0}

\section*{\normalsize 1 \ Introduction}

One of the outstanding problems of modern cosmology is the
understanding of structure formation in the universe \cite{peebles93}. 
Visible matter at astronomical scales appears organized in a hierarchy of
galaxies, cluster and supercluster of galaxies, which tend to be
found in filamentary aggregates or two-dimensional sheets that
encompass large regions with much lower density of matter and
structures. These regions, called voids, 
occupy a large fraction of the observed universe, and are also organized in a 
hierarchy \cite{einasto1,einasto2}. 
The theoretical investigation of void formation
is receiving increasingly attention \cite{peebles01}. 

There is general agreement that the galaxy two point correlation
function is scale invariant (self-similar), at least at not too 
large scales, decaying as
a power of the distance, $1/r^\gamma$. 
The exponent $\gamma$ as well as the scale
of homogeneity is still controversial. Pietronero and coworkers
found $\gamma\approx 1$ and claim that the power law is obeyed up to
the deepest distances \cite{pietronero}. Other authors gave a different
exponent ($\gamma\approx 1.8$) and support the existence of an observed 
scale of homogeneity 
\cite{nonfractal:davis,nonfractal:peebles,nonfractal:guzzo}. 
Self-similar behaviour has also been found in the interstellar medium 
\cite{ism:larson,ism:hetem,ism:kramer,ism:stutzki}.

Although certainly the dynamics must be very important in order to explain
these facts, and many physical effects might play a prominent role, 
it is well possible that many aspects of the observed structures may
be understood in terms of the equilibrium states of self-gravitating
matter alone, as claimed by the authors of Refs. \cite{dvsc,pm}.
Indeed, it has been tried to apply thermodynamics to astrophysical systems
since a long time (see for instance Ref. \cite{saslaw}).
In this paper, we will analyze the general phase structure of a 
non-relativistic self-gravitating system at thermal equilibrium. 
The results are general and can be applied to any such system that can be
considered to be in these conditions. We will not be concerned here with the 
very interesting questions of whether thermal equilibrium  can be reached in
such systems, or the way it is attained. We just assume that the
self-gravitating system is at thermal equilibrium.

Let us consider a system of $N$ classical particles of mass $m$ confined
in a region of volume $V$ and
interacting each other via the Newtonian gravitational potential.
Its hamiltonian reads
\begin{equation}
{\cal H}\;=\;\frac{1}{2m}\,\sum_{i=1}^N \pvec_i^2-\:\frac{G m^2}{2}\,
\sum_{i\neq j} \frac{1}{|\rvec_i-\rvec_j|} \, . \label{hamil}
\end{equation}
The thermodynamics of such a system is ill defined: the entropy  
does not exist due to the singularity of the potential at short
distances \cite{pad,eduardo}. 
Furthermore, even if one modifies the potential 
to remove the short distance singularity, the usual thermodynamical
limit does not exist, since the thermodynamic potentials are not extensive,
due to the long range gravitational force. 
In such cases, the microcanonical specific
heat can become negative and different statistical
ensembles are not equivalent (see Refs.
\cite{gross,lynden-bell,thirring,stahl,torcini,chav:som,youngkins,ispolatov,chavanis,chav:isp}). 
In particular, the grand canonical ensemble
is dominated by completely collapsed configurations, whatever the chemical
potential.
To use the ordinary thermodynamical tools, the potential must also be
modified at long distances\footnote{It has been proposed
that a kind of thermodynamical limit for systems with potentials decaying 
as $1/r$ could be taken by considering the so called
dilute regime \cite{devega}, but it can be shown that this statement 
cannot hold \cite{victor2}.}.

The thermodynamics of self-gravitating system has been studied since a
long time by confining a finite system on a finite box and using
mean field theory.
The approach developed in this paper is different: we will
describe the thermodynamical
properties of a self-gravitating system as the limiting case of a
family of well behaved, short range systems, the interactions of which
decay with distance more and more slowly, thus resembling more and more
the newtonian $1/r$ potential. For these systems the usual thermodynamical
limit, which is considered in this paper, does exist. 
To carry out this investigation, we will take advantage of the fact that
the statistics of a self-gravitating ensemble of particles can be related 
to an euclidean field theory of a single scalar field \cite{dvsc,pm},
in a similar way to the relation between the Coulomb gas and the
Sine-Gordon field 
theory \cite{coulomb:froelich,coulomb:polyakov,coulomb:samuel}.
Hence, the remaining of the paper will rely on the techniques of 
euclidean field theory.

\section*{\normalsize 2 \ Field theoretical description of a self-gravitating 
system}

The statistical mechanics of a self-gravitating system can be studied 
in the usual way if the gravitational potential is regularized at short 
and long distances. 
Let us choose as a regularized gravitational potential, 
denoted by $\rpot(r)$,
an attractive Yukawa potential, with range $1/m_0$, endowed  
with a hard core of size $r_0$ at short distances: $\rpot(r)=
-Gm^2\mathrm e^{-m_0 r}/r$ for $r>r_0$
and $\rpot(r)=\infty$ for $r<r_0$. 
After integrating out the momenta, the grand canonical partition 
function for chemical potential $\mu$
and temperature $T$ can be written as
\begin{equation}
{\cal Z}_{\textrm{\tiny GC}}=\sum_N \,\mathrm e^{\mu N}\, 
\left(\frac{m a^2}{2\pi\beta\hbar^2}\right)^{3N/2}
\bar{\cal Z}_N\, ,
\label{gcpf}
\end{equation}
where  
$\beta=1/k_BT$, $a$ is a constant with units of length, to
be specified below, and 
\begin{equation}
\bar{\cal Z}_N=\frac{a^{-3N}}{N!}
\int\prod_{i=1}^Nd^3r_i\,\exp\{-\frac{\beta}{2}
\sum_{i\neq j}\rpot(|\rvec_i-\rvec_j|)\} \, .
\end{equation}

The partition function $\bar{\cal Z}_N$ can be approximated by
dividing the space volume in $\Lambda=V/a^3$ cells of size $a$ (the
unit length introduced above),  and replacing 
the integrals by appropriate Riemann sums, which can be reorganized as a
sum over cell occupation numbers $S_i$, with the index $i$ running from
$1$ to the number of cells, $\Lambda$. If $a$ is of the order of the
hard core size, each cell can be either void or occupied by one particle:
$S_i=0,1$. This is similar to consider self-gravitating 
fermions \cite{chavanis}. We get
\begin{equation}
\bar{\cal Z}_N \,=\, \sum_{S_1=0}^1\ldots\sum_{S_{\Lambda}=0}^1
\,\exp\left\{\frac{b^2}{2}\left[\sum_{n,n'}\,
U_{\textrm{\tiny L}}(n-n') 
S_n S_{n'}-U_{\textrm{\tiny L}}(0) N \right]\right\} \:
\delta(\sum_n S_n - N) \, ,
\label{discpf}
\end{equation}

\noindent
where $b^2=4 \pi\beta G m^2$,  \mbox{$U_{\textrm{\tiny L}}(n-n')$} 
is a proper lattice version of 
$-\rpot(\rvec-\rvec')/(4\pi G m^2)$,
the factor $1/N!$ has been canceled due to the distinguishability
of classical
particles, and the Kronecker delta takes into account that the number
of particles is fixed \cite{victor}.

Since the Yukawa potential is the Green function of
$-\nabla^2+m_0^2$,
we can choose its lattice version as the lattice Green function
of $-\nabla_{\textrm{\tiny L}}^2+m_0^2$, where 
$\nabla_{\textrm{\tiny L}}^2$ is a discretization of the 
laplacian:
\begin{equation}
U_{\textrm{\tiny L}}(n-n')\;=\;\frac{1}{a^3}\,
\left(\frac{1}{-\nabla^2_{\textrm{\tiny L}}+m_0^2}\right)_{n,n'}\, .
\label{latpot}
\end{equation}
Using (\ref{latpot}) and the Hubbard-Stratonovich
formula, we can write the following identity:
\begin{eqnarray}
&&\exp\left\{\frac{b^2}{2}\sum_{n,n'}\,U_{\textrm{\tiny L}}(n-n') S_n
S_{n'}\right\} \;= \nonumber \\
&& \;\;\;\;\;\; 
C\,\int\,[d\psi]\,\exp\left\{-\frac{1}{2}\,\sum_{n,n'}\,a^3\,
\psi_n\,\left(-\nabla^2_{\textrm{\tiny L}}+m_0^2\right)_{nn'}\,\psi_{n'}\:+\:
b\,\sum_n\psi_n S_n\right\} \, ,
\label{hs}
\end{eqnarray}

\noindent
where $C$ is a number independent of $S_n$. Plugging Eq. (\ref{hs}) in
(\ref{discpf}) and the resulting equation for $\bar{\cal Z}_N$ in the
expression for $\cal{Z}_{\textrm{\tiny GC}}$, and performing the summation 
over $S_i$, we can write 
the grand canonical partition function of
the (regularized) self-gravitating system in terms of a local euclidean field 
theory with a single scalar field $\psi_n$:
\begin{equation}
{\cal Z}_{\textrm{\tiny GC}}=C\int[d\psi]\exp[-{\cal S}]\, ,
\end{equation}
the action of which is given by
\begin{equation}
{\cal S}=\frac{1}{2}\sum_{nn'}a^3
\psi_n{\cal D}_{nn'}\psi_{n'}-
\sum_n \ln\left(1+g \mathrm e^{b\psi_n}\right)\, ,
\label{firstaction}
\end{equation}
where $g=\mathrm e^{\mu}\mathrm e^{-\frac{b^2}{2}U_{\textrm{\tiny L}}(0)}
(\frac{m a^2}{2\pi\beta\hbar^2})^{3/2}$ and
${\cal D}=-\nabla^2_{\textrm{\tiny L}}+m_0^2$.
Notice the unusual form of the action: the
interaction term, $\ln(1+g\mathrm e^{b\psi})$, is unbounded from
below. It behaves as $-b\psi$ for $\psi\rightarrow\infty$. The action, 
however, is bounded from below if $m^2_0>0$. This reflects the fact
that the grand canonical ensemble does not exist for non-extensive
(long range and/or purely attractive) systems. The infrared and
ultraviolet cutoffs, $m_0$ and $a$ respectively, make the system
short ranged and repulsive at short distances (hard core of size
of the order of $a$), and thermodynamics is well defined. 
In the naive continuum limit, in which
$a\rightarrow 0$ ($g\rightarrow 0$)
and the relevant field configurations are smooth, 
we recover the action of Ref. \cite{dvsc}. This continuum action is not 
bounded from below and hence the corresponding functional integral 
diverges. Thus, the results of Ref. \cite{dvsc} are at most formal
and the conclusions different from ours. The regulators, $a$ and $m_0$,
play an essential role, although the conclusions, as we will see,
are independent of them (provided they are small enough).

The action contains four parameters: the lattice spacing, $a$, which
obviously has dimension of length; the inverse of the potential range,
$m_0$, which has dimension of $(\textrm{length})^{-1}$; $b$, with
dimension of $(\textrm{length})^{1/2}$; and $g$, which is
dimensionless. The field is canonically normalized, so that its
dimension is $(\textrm{length})^{-1/2}$.
It is convenient to redefine the field such that $\phi_n=b\psi_n+\ln g$,
and to work in terms of the dimensionless 
quantities $\kappa=a/b^2$, $\mass=a^2 m_0^2$, and 
$h=1/2+\kappa\mass\ln g$. Ignoring constant terms, the action reads
\begin{equation}
{\cal S}=\frac{\kappa}{2}\sum_{nn'}
\phi_n \overline{\cal D}_{nn'}\phi_{n'}
-\sum_n \ln\cosh\frac{\phi_n}{2}
-h\,\sum_n\phi_n\, ,
\label{action}
\end{equation}
where $\overline{\cal D}=-\lapl+\mass$.
An action of the same form, with $\mass=0$, has been obtained in 
Ref. \cite{pm} in a similar way. In that case, however, 
the parameter $h$ has a different meaning.
Notice that for $h=0$ the action (\ref{action}) has the symmetry 
$\phi_n\rightarrow -\phi_n$. Since the term linear in $\phi_n$ 
breaks this 
symmetry, we will have $\langle\phi\rangle>0$ for $h>0$ and
$\langle\phi\rangle<0$ for $h<0$. At $h=0$ we will have
$\langle\phi\rangle=0$ unless the symmetry is spontaneously broken.

The average of particle and energy densities, $\rho$ and $\epsilon$, 
can be obtained as 
derivatives of the grand canonical partition function. This allows
to express them in terms of the field $\phi$ in the following way:
\begin{eqnarray}
\rho &=&
\frac{1}{V}\sum_n\frac{\mathrm e^{\phi_n}}{1+\mathrm e^{\phi_n}}\, , 
\label{density} \\*
\epsilon &=& 
k_B T\left[\frac{3}{2}\rho-\frac{1}{2}
\frac{1}{V}\sum_n[\phi_n-\ln g-\kappa^{-1}\overline{\mathcal{D}}^{-1}_{nn}]
\frac{\mathrm e^{\phi_n}}{1+\mathrm e^{\phi_n}}
\right] \, .
\label{energy}
\end{eqnarray} 
One can easily identify the first contribution to the energy in 
Eq.~(\ref{energy}) as the kinetic energy and the second one as the
potential energy. The later has the form 
$\sum_n (\phi_n-\ln g-\kappa^{-1}\overline{\mathcal{D}}^{-1}_{nn})\rho_n$, 
where $\rho_n=\mathrm e^{\phi_n}/(1+\mathrm e^{\phi_n})$ is the local 
particle density, what implies that the field
$\phi_n$ represents, up to an additive constant,
$\ln g+\kappa^{-1}\overline{\mathcal{D}}^{-1}_{nn}$,
the local gravitational potential\footnote{Note that 
$\overline{\mathcal{D}}^{-1}_{nn}$ is independent of $n$ due to the 
translational invariance of the Laplacian.}. 

After some algebraic manipulations, which are outlined in the appendix
since they are not completely straightforward, we can write the following 
exact equation for the average energy, which has the expected form:
\begin{equation}
\epsilon=k_B T\left[\frac{3}{2}\rho-\frac{1}{2}\frac{1}{V}\sum_n\sum_{r\neq 0}
\kappa^{-1}\langle\rho_n\overline{\mathcal{D}}^{-1}_{n,n+r}\rho_{n+r}\rangle
\right]\, . 
\label{avenergy}
\end{equation}
In the above equation $\epsilon$ and $\rho$ stand for the average energy and
density, respectively. Remember that 
$\overline{\mathcal{D}}^{-1}_{n,n+r}\approx \mathrm e^{-\overline{m}_0 r}/r$.
Notices that the fact that the term with $r=0$ is
excluded from the sum over $r$ in Eq.~(\ref{avenergy}) implies that the
contact term does not contribute to the energy. This, and the bounds
$0<\rho_n<1$, are manifestations of the particle hard core.

\section*{\normalsize 3 \ Phase diagram}

Perturbation theory in euclidean field theory starts by identifying
the minimum of the classical action and assuming that the relevant
field configurations are small fluctuations around this minimum. 
For the action (\ref{action}) this is a good approximation if $\kappa$
is large. We shall argue below that it is indeed 
a good approximation whatever $\kappa$ if $\mass$ is small.

The minimum of the action obviously corresponds to the constant field that 
minimizes the classical potential
\begin{equation}
{\cal U}\;=\;\frac{\kappa\mass}{2}\phi^2\:-\:
\ln\cosh\frac{\phi}{2}\:-\:h\phi \, ,
\label{claspot}
\end{equation}
and, therefore, satisfies the equation
\begin{equation}
\kappa\mass\phi\;=\;\frac{1}{2}\,\tanh\frac{\phi}{2}\:+\:h \, .
\label{minimum}
\end{equation}

\begin{figure}[t!]
\centerline{\includegraphics*[width=2.5in,angle=90]{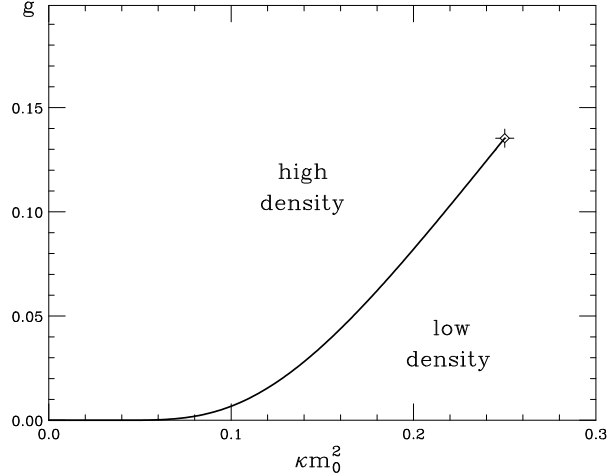}}
\caption{Phase diagram in the $(\kappa\mass,g)$ plane. The solid
line is a first order transition that ends at the critical point.}
\label{fig:phd}
\end{figure} 

The above equation has one, two or three solutions, depending on
the values of $\kappa\mass$ and $h$. When it has one solution, it 
correspond to the global minimum; if there are two solutions,
one is the global minimum and the other one a point of inflection;
three solutions correspond to a local maximum, a local minimum,
and the global minimum. In the last case, the minima can only be 
degenerate if $h=0$, and then symmetry implies that 
Eq.~(\ref{minimum}) has either one or three solutions: 
for $\kappa\mass>1/4$ the only solution is $\phi=0$, 
while for $\kappa\mass<1/4$ we have two
solutions, $\phi=\pm\phi_0\neq 0$ (with $\phi_0$ positive), 
which are the two degenerate
global minima, besides the symmetric solution $\phi=0$, which is a
local maximum. We will denote the two global minima by
$\phi_h=+\phi_0$ and $\phi_l=-\phi_0$.
The densities of each phase are respectively
\begin{equation}
\rho_l=\frac{1}{a^3}\,\frac{\mathrm e^{\phi_l}}{1+\mathrm e^{\phi_l}}\, ,
\hspace{1truecm} 
\rho_h=\frac{1}{a^3}\,\frac{\mathrm e^{\phi_h}}{1+\mathrm e^{\phi_h}}\, .
\end{equation}
Hence, the solutions $\phi_l$ and $\phi_h$ describe phases 
of low and high density respectively.

The phase diagram in the plane $(\kappa\mass,g)$, at the classical 
level, is displayed in figure~\ref{fig:phd}. For $\kappa\mass>1/4$, the
transition between the low (small $g$) and high (large $g$)
density regimes is smooth. For
$\kappa\mass<1/4$, the low density phase is separated from the
high density phase by a first order transition which takes place
at $\gfo=\exp[-1/(2\kappa\mass)]$ (i.e., $h=0$). 
This first order line ends at
the critical point $\kappa_c\mass=1/4$, 
$g_c=1/\mathrm e^2$ ($h=0$). The order parameters,
$\delta\phi=\phi_h-\phi_l$ and
$\delta\rho=\rho_h-\rho_l$,
vanish at the critical point with the classical (mean field) 
exponent $1/2$. This mean field critical behavior is a consequence of the
classical approximation and could be modified by the neglected 
fluctuations.

For small $\kappa\mass$ and $h=0$, Eq.~(\ref{minimum}) gives
$\phi_h\approx 1/(2\kappa\mass)$,
$a^3 \rho_h\approx 1-\mathrm e^{-1/(2\kappa\mass)}$,
$\phi_l\approx -1/(2\kappa\mass)$, and
$a^3 \rho_l\approx\mathrm e^{-1/(2\kappa\mass)}$.
Thus, the low density phase is very dilute, while the high density 
phase is extremely dense. Most of the densities (including 
presumably those of physical systems) are between $\rho_l$
and $\rho_h$, and they correspond, therefore, to the phase
coexistence region. 

\section*{\normalsize 4 \ Critical behavior and correlations}

The corrections to the classical approximation can be obtained
perturbatively, by expanding the action around the corresponding
minimum. For each phase we write
$\phi_n=\phi_h+\varphi_n$ and $\phi_n=\phi_l+\varphi_n$
respectively, and, ignoring again constant terms, the corresponding 
actions can be written as

\begin{equation}
{\cal S}\;=\;\frac{\kappa}{2}\,\sum_{nn'}\,\varphi_n
\overline{\cal D}_{nn'}\varphi_{n'}
\:\pm\:\lambda\,\sum_n\varphi_n 
\:-\:\sum_n\,\ln\left[1+\lambda (\mathrm e^{\pm\varphi_n}-1)\right]
\, ,  
\label{action_min}
\end{equation}
where the signs $+$ and $-$ correspond to the low and high density
phases respectively, and
$\lambda=1-a^3\rho_h=a^3\rho_l$.

When $\kappa\mass$ is very small, $\lambda$ is much smaller and the
actions for either phase (\ref{action_min}) are 
very close to the gaussian critical point, $\lambda=0$. Therefore, 
the classical
approximation we are using is very good. In addition, we have a very
interesting situation: \textit{a very sharp first order transition that
separates two phases which, in turn, are close to criticality}.
Hence, the actions (\ref{action_min}) produce
a very large correlation length and critical (self-similar) behavior
over a vast range of scales. The question is to which universality
class corresponds such critical behavior. In three dimensions, the
gaussian fixed point is infrared unstable under perturbations of
relevant operators. This means that the critical behavior of actions
defined in the neighborhood of the gaussian fixed point can be 
governed by a different (non-gaussian) fixed point. For this 
to happen, the couplings of the relevant operators of dimension
larger than one must be of the order of $\kappa\mass$.
In our case $\lambda\ll\kappa\mass$ and we are in the 
opposite case: the actions~(\ref{action_min}) lie very close to the
renormalized trajectory of the gaussian fixed point, given by
$\lambda=0$, $\kappa\mass>0$. Hence, the critical behavior is governed
by the gaussian fixed point, and thus belongs to the mean field
universality class. 

The correlations can be computed to leading order in $\lambda$
with good accuracy. For the field we get
\mbox{$\langle\phi_{\rvec}\phi_{\rvec'}\rangle
-\langle\phi_{\rvec}\rangle\langle\phi_{\rvec'}\rangle
=G(\rvec-\rvec')$}, with
\begin{equation}
G(\rvec-\rvec')\;=\;\int\,\frac{d^3p}{(2\pi)^3}\,
\frac{\mathrm e^{\mathrm i \pvec (\rvec-\rvec')}}{p^2+\kappa\mass}\, ,
\end{equation}
where we have used the continuum expression since the lattice spacing
is much smaller than the correlation length, $\xi=a/(\kappa\mass)^{1/2}$. 
At large distances, $G(\rvec)$ decays exponentially as 
$\exp(-|\rvec|/\xi)$, but in a wide range of distances, $a\ll|\rvec|\ll\xi$,
it is approximately self-similar, $G(\rvec)\sim 1/|\rvec|$. 

The correlations of the density (\ref{density}), 
$\Gamma(\rvec-\rvec')$, have the same long distance behavior as
the correlations of the field.
To leading order in $\lambda$ we have:
\begin{equation}
\Gamma(\rvec-\rvec')\;=\;\lambda^2\,\mathrm e^{G(0)}\,
\left[\mathrm e^{G(\rvec-\rvec')}-1\right]\, .
\end{equation}
For large distances $\Gamma(\rvec)\sim G(\rvec)$, and therefore
the density correlations behave as $1/|\rvec|$ for $a\ll|\rvec|\ll\xi$.
Hence, the present approach to self-gravitating systems predicts 
that density correlations decay as a power 
law with exponent $\gamma=1$ over a vast range of scales. Since $\xi$
is proportional to the assumed range of the gravitational interaction,
it is well possible that correlations be self-similar at any
observable scale.

It is worthwhile stressing that the above analysis refers to the 
quasi-critical behaviour
of each of the two phases as $\kappa\mass\rightarrow 0$. There is a
true critical point at $\kappa_c\mass=1/4$ and $g=1/\mathrm e^2$
(or $\rho_c=1/2$). In this case the parameters are not small
and a non-perturbative analysis is required in order to investigate the
nature of its universality class. This critical point might be
relevant at very high temperatures, when $\kappa\mass$ is of order one.

A similar analysis can be made for small $\kappa$ (strong coupling
regime). The symmetry is broken at the classical level
if $\kappa\mass<1/4$. For small $\kappa$ and small $\mass$
short wavelength fluctuations cannot induce tunneling between the two
minima since it would imply huge fluctuations of the action, of the order 
$1/\kappa\overline{m}_0^4$. Likewise, long wavelength fluctuations 
cannot induce tunneling since the barrier is too high,
of the order of $1/(8\kappa\mass)$.
The two minima describe phases of low and high density. 
The correlation length on each phase
is of the order of $1/\mass$. Hence, the conclusions are the same
as in the weak coupling regime: an abrupt first order transition
separates the low and high energy phases, and each phase present
critical behaviour of mean field type.

\section*{\normalsize 5 \ Finite volume effects: the Lane-Emden equation}

It might seem paradoxical that the mean field solution is homogeneous,
since it is well known that the usual mean field solutions of self-gravitating 
systems present density profiles that depend on the distance to the center
of the system. There is
a simple explanation: since we are looking for the mean field
solution of a short ranged translational invariant system in the
infinite volume limit it is natural that it be homogeneous. 
Had we looked for the minimum of the action (\ref{firstaction}) on a 
finite box of size $L$ we would have
obtained the equation
\begin{equation}
\sum_{n'}(-\nabla_L^2)_{nn'} \psi_{n'}\:+\:m_0^2\psi_n\:-\:
\frac{b g e^{b\psi_n}}{1+g e^{b\psi_n}} \;=\;0\, .
\label{finite}
\end{equation}
On a finite volume, the walls break translational invariance and 
the solution of Eq.~(\ref{finite}) will not be homogeneous, due to the
boundary conditions imposed to the operator $(-\nabla_L^2)_{nn'}$.
Since this operator is a discretization of the Laplacian, if
$L \ll 1/m_0$, Eq.~(\ref{finite}) is similar to the isothermal Lane-Emden 
equation \cite{emden,chandra}
(if $m_0=0$, to leading order in $g$
we will have exactly the discretized Lane-Emden equation, and the 
corrections in $g$ are due to the cut-off $a$),
and we will get a solution similar
to the known profiles of self-gravitating systems in the mean field
approximation. 
On the other hand, if $L\gg 1/m_0$, the spatial dependence of the solution
will be washed out
and we will have the reported constant solutions. However,
the two solutions correspond to extremely high and extremely low densities, 
respectively. The system will only be homogeneous at such extreme densities.
For intermediate densities, inhomogeneities will necessarily develop: the
system will be in a mixture of high and low density domains.

This instability as the volume increases had been notice long ago by
Antonov, who found that the solutions of the Lane-Emden equation 
ceases to be (local) maxima of the entropy if the size of the box
is larger than $0.335 G M^2/(-E)$, where $M$ is the total mass and $E<0$ the
energy, and the system collapses \cite{antonov}. This was called
the gravothermal catastrophe in Ref. \cite{wood}.

\section*{\normalsize 6 \ Final remarks}

In thermal equilibrium, a self-gravitating system is made up
of domains of low and high density (voids and clusters).
The distribution of domains is a dynamical question that
has to do with the way equilibrium is reached.
The correlations within each domain are self-similar on a vast range
of scales, decaying as $1/r$. The transition to
homogeneity\footnote{The scale at which the density 
fluctuations cease to be self-similar.} \cite{gaite}  
would take place on scales comparable to the range
of the gravitational interaction, which may be larger than the
deepest observed distances. A similar behavior (first order phase
transition and self-similar correlations) was found for the
two-dimensional self-gravitating system by using techniques of
conformal field theory \cite{abdalla}.

It is straightforward to take into account the cosmic
expansion in a simplified way, as in Refs. \cite{pm,dvsc}, by introducing
comoving coordinates $\xvec$ in (\ref{hamil}), such that the physical
coordinates are $\rvec=R(t)\xvec$, where $R(t)$ is the scale factor.
As intuitively expected, this is equivalent to rescale the lattice spacing:
$a(t)=R(t) a$ and, consequently, we have the following rescaling of 
parameters: 
$\kappa\rightarrow R(t)\kappa$ and $\mass\rightarrow R^2(t)\mass$.
This variation of the parameters implies that the difference between
the densities of the two phases decreases. This is not surprising:
the expansion of the universe acts as a pressure that competes with
the tendency of gravity towards collapse.

It is remarkable that the picture of the self-gravitating system 
devised in this work is strikingly similar to the observed universe, 
despite it is not at thermal equilibrium. The scaling exponent
of the the correlation function of the density, $\gamma=1$,
however, is far from the
widely accepted exponent of the galaxy correlation function,
$\gamma\approx 1.8$ for distances between 0.2 and 20 Mpc \cite{peebles93}. 
As mentioned in the introduction, in the last years
there has been a strong debate about the validity of this result.
Pietronero and his group analyzed the data 
from a different perspective and claimed that the exponent of the
galaxy correlation function is $\gamma \approx 1$, and that the
self-similar behavior extends up to the deepest explored distances.
The controversy seems not resolved, although most astrophysicist 
believe the earlier result, $\gamma\approx 1.8$. 
This is theoretically supported by the thermodynamic 
arguments given in \cite{saslaw}. These arguments, however, are
somehow heuristic and their validity may be questioned. Indeed, using 
Eq.~(\ref{avenergy}) as starting point,
the same arguments can be applied step by step to the formulation of 
the self-gravitating system given in this work, leading to the 
same predictions of Saslaw's book. However, the results of this work, 
based on a rather rigorous treatment of the self-gravitating system, 
are completely different.

Hence, the conclusion of this paper is clear: either Pietronero and 
coworkers are right,
and $\gamma\approx 1$, or the thermodynamical approach is not valid
to describe the universe at such scales. The later possibility cannot 
be discarded, since it is difficult to argue that the universe is at 
thermal equilibrium, and 
dynamical effects may play a prominent role in the behavior of
the galaxy correlation function.
In any case, the results obtained in this paper need not be applied to
the universe as a whole, or to the large scale structure of it,
but to any self-gravitating system that are close to
thermal equilibrium. The interstellar gas, where self-similar behavior
has also been observed, might be an instance. In this case the scaling 
exponent of the density correlations is compatible with $\gamma=1$, but 
with large uncertainties \cite{dvsc}.

\vspace{0.25truecm}

\noindent
\textit{Acknowledgments}

I thank J.L. Alonso, J.L. Cort\'es, and A. Galante for useful
discussions. This work received financial support from Ministerio de
Ciencia y Tecnolog\'{\i}a (Spain) under the Ram\'on y Cajal program,
and from CICyT (Spain) under the project FPA 2000--1252.

\section*{\normalsize Appendix}

Let us outline in this appendix the derivation of equation~(\ref{avenergy}),
which is not completely straightforward. The field $\phi_n$ can be obtained
from the derivative of the action as
\begin{equation}
\phi_n=\kappa^{-1}\sum_{n'}\overline{\mathcal{D}}^{-1}_{nn'}\rho_{n'}+
\kappa^{-1}\sum_{n'}\overline{\mathcal{D}}^{-1}_{nn'}
\frac{\partial \mathcal{S}}{\partial\phi_{n'}}
-\frac{1/2-h}{\kappa\mass}\, ,
\end{equation}
where we used $\sum_{n'}\overline{\mathcal{D}}^{-1}_{nn'}=1/\mass$.
Now insert the above expression for $\phi_n$ in 
equation~(\ref{energy})\footnote{Not in the term 
$\rho_n=\mathrm e^{\phi_n}/(1+\mathrm e^{\phi_n})$.}.
The term proportional to $(1/2-h)/\kappa\mass$ cancels the $\ln g$ term
of Eq.~(\ref{energy}).
Averaging over the thermal fluctuations the resulting expression for 
$\epsilon$, we are
led to compute $\langle\rho_n\partial\mathcal{S}/\partial\phi_{n'}\rangle$,
which can be written as
\begin{equation}
\left\langle\rho_n\frac{\partial\mathcal{S}}{\partial\phi_{n'}}\right\rangle
=-\frac{1}{\mathcal{Z}}\int [d\phi] \rho_n\frac{\partial}{\partial\phi_{n'}}
\mathrm e^{-\mathcal{S}}\, .
\end{equation}
Integrating by parts and taking into account that 
$\partial\rho_n/\partial\phi_{n'}=\delta_{nn'} \rho_n (1-\rho_n)$\, ,
we get
\begin{equation}
\left\langle\rho_n\frac{\partial\mathcal{S}}{\partial\phi_{n'}}\right\rangle
=\delta_{nn'} \langle\rho_n (1-\rho_n)\rangle\, .
\end{equation}
The term linear in $\rho_n$ cancels the term proportional to 
$\kappa^{-1}\overline{\mathcal{D}}^{-1}_{nn}$  of the averaged 
Eq.~(\ref{energy}). Hence, we obtain for the averaged energy
\begin{equation}
\epsilon=k_B T\left[\frac{3}{2}\rho-\frac{1}{2}\frac{1}{V}\sum_{nn'}
\kappa^{-1}\langle\rho_n\overline{\mathcal{D}}^{-1}_{nn'}\rho_{n'}\rangle
+\frac{1}{2}\frac{1}{V}\sum_{n}\kappa^{-1}
\langle\rho_n\overline{\mathcal{D}}^{-1}_{nn}\rho_{n}\rangle\right]
\, .
\end{equation}
It suffices to write $n'=n+r$ to realize that the above equation is 
Eq.~(\ref{avenergy}).

\end{document}